\documentclass{Interspeech2024}
\usepackage{amsmath,graphicx,url}
\usepackage{subcaption}

\interspeechcameraready

\title{\texttt{AUREXA-SE}: Audio-Visual Unified Representation Exchange Architecture with Cross-Attention and Squeezeformer for Speech Enhancement}
\name[affiliation={1}]{M.}{Sajid}
\name[affiliation={1}]{Deepanshu}{Gupta}
\name[affiliation={1}]{Yash}{Modi}
\name[affiliation={1}]{Sanskriti}{Jain}
\name[affiliation={1}]{Harshith}{Jai Surya Ganji}
\name[affiliation={1}]{~~~~~~~~~~~~~~~~~~~~~~~~~~~~A.}{Rahaman}
\name[affiliation={1}]{Harshvardhan}{Choudhary}
\name[affiliation={2}]{Nasir}{Saleem}
\name[affiliation={2}]{Amir}{Hussain}
\name[affiliation={1}]{M.}{Tanveer}

\address{
  $^1$Indian Institute of Technology Indore, Simrol, Indore, 453552, India\\
  $^2$School of Computing, Edinburgh Napier University, EH11 4BN, Edinburgh, United Kingdom}
\email{A.Hussain@napier.ac.uk, mtanveer@iiti.ac.in}


\begin{document}
\maketitle

\begin{abstract}
In this paper, we propose \texttt{AUREXA-SE} (Audio-Visual Unified Representation Exchange Architecture with Cross-Attention and Squeezeformer for Speech Enhancement), a progressive bimodal framework tailored for audio-visual speech enhancement (AVSE). \texttt{AUREXA-SE} jointly leverages raw audio waveforms and visual cues by employing a U-Net–based 1D convolutional encoder for audio and a Swin Transformer V2 for efficient and expressive visual feature extraction. Central to the architecture is a novel bidirectional cross-attention mechanism, which facilitates deep contextual fusion between modalities, enabling rich and complementary representation learning. To capture temporal dependencies within the fused embeddings, a stack of lightweight Squeezeformer blocks combining convolutional and attention modules is introduced. The enhanced embeddings are then decoded via a U-Net–style decoder for direct waveform reconstruction, ensuring perceptually consistent and intelligible speech output. Experimental evaluations demonstrate the effectiveness of \texttt{AUREXA-SE}, achieving significant performance improvements over noisy baselines, with STOI of 0.516, PESQ of 1.323, and SI-SDR of -4.322~dB.
The source code of \texttt{AUREXA-SE} is available at \href{https://github.com/mtanveer1/AVSEC-4-Challenge-2025}{https://github.com/mtanveer1/AVSEC-4-Challenge-2025}.
\end{abstract}

\section{Introduction}
Speech is central to human communication, enabling information exchange and social connection. However, its intelligibility often deteriorates in noisy environments, making effective communication challenging without accurate interpretation. Enhancing speech clarity through audio-visual modeling is therefore vital for building robust human-system interfaces in practical environments \cite{anwary2024target}. This has led to the development of the field of speech enhancement (SE), which aims to improve both the coherence and quality of speech \cite{tanveer2023ensemble, yechuri2025speech}. Traditional SE methods primarily relied on time-frequency (TF) domain processing using Convolutional Neural Networks (CNNs) \cite{pandey2018new} or Recurrent Neural Networks (RNNs) \cite{pandey2022self}. A notable milestone was the Convolutional Recurrent Network (CRN) \cite{tan2018convolutional}, which integrated a convolutional encoder-decoder with Long Short-Term Memory (LSTM) \cite{6795963} units to precisely characterise variation in speech \cite{jain2024lstmse}. Later innovations like the Deep Complex Convolution Recurrent Network (DCCRN) \cite{hu2020dccrn} extended these ideas by processing complex-valued spectrograms, which led to significant gains in both objective and subjective speech quality. Following this trend, end-to-end waveform-based models using Generative Adversarial Networks (GANs) \cite{10890549} have shown impressive adaptability across diverse speakers and noisy environments.

Despite this significant progress in audio-only speech enhancement using deep learning, such approaches remain limited in highly noisy or acoustically challenging environments. These methods often struggle when the signal-to-noise ratio (SNR) is low or when the noise characteristics overlap with speech. Crucially, they lack access to the complementary contextual information that humans naturally rely on during communication, as demonstrated by phenomena like the McGurk effect \cite{tiippana2014mcgurk}. Hence, recent research has increasingly turned to AVSE, where visual cues such as lip movements provide noise-agnostic information to aid speech recovery.

With the aid of temporally aligned visual information, audio-visual models are able to recover speech more effectively in challenging environments. By leveraging state-of-the-art (SOTA) models and techniques, multimodal architectures are able to capture richer contextual patterns, significantly improving speech enhancement performance. In essence, a unified framework that synergizes recent advances across both audio and visual domains holds the potential to address limitations that traditional approaches have struggled to overcome.

Motivated by this insight, we propose \texttt{AUREXA-SE}, a novel AVSE framework developed for the COG-MHEAR AVSE Challenge. It features a dual-stream architecture with a U-Net-based audio encoder \cite{luo2019conv} and a Swin Transformer V2 visual encoder \cite{liu2022swin}, each processing their modality in parallel. The extracted features are fused using bi-directional cross-attention \cite{lin2022cat}, refined via Squeezeformer-based temporal modelling \cite{kim2022squeezeformer}, and decoded using a U-Net–style waveform decoder \cite{stoller2018wave} to generate clean speech. Together, these components form a unified, cross-modal architecture that combines the strengths of both audio and visual inputs while introducing key innovations, such as raw waveform encoding, spatially rich visual processing, and bi-directional cross-modal fusion. This fusion of best practices is aimed at delivering robust speech clarity in noisy environments while maintaining computational efficiency and scalability.

Our framework achieves state-of-the-art speech enhancement within a focused training budget of just 50 hours—20 epochs at 2.5 hours each (348,660 steps)—outperforming models that require significantly longer training schedules. With an expanded architecture of 54.2M parameters (up from the baseline’s 22.2M), we deliver notable quality improvements. Despite a modest increase in inference time (40 minutes vs. 25 minutes), the gains clearly outweigh the trade-off, showcasing the efficiency and effectiveness of our design.

In the following section, we will provide a comprehensive overview of related works in both audio-only and audio-visual speech enhancement, further contextualising the challenges in the field and highlighting how \texttt{AUREXA-SE}'s innovative design directly addresses these limitations.

\section{Motivation and Contribution}
SE has undergone a significant transformation, moving beyond traditional audio-only methods to embrace multimodal approaches that leverage visual information. This paradigm shift is motivated by the recognition that visual cues such as lip movements and facial expressions provide temporally aligned and noise-resilient context. By enriching speech representations and resolving acoustic ambiguities, visual input has catalyzed a growing interest and rapid advancement in AVSE research.

The progress in AVSE has given rise to several notable architectures. RecognAVSE \cite{manesco2024recognavse} innovatively combines a Separable 3D CNN \cite{yin2023convolutional} for efficient video encoding with a DCU-Net \cite{sun2021funnel} audio encoder. Meanwhile, LSTMSE-Net \cite{jain2024lstmse} leverages an LSTM-based network to process concatenated audio-visual features, consistently outperforming recent challenge baselines. More recently, DAVSE \cite{chen2024davse} introduced a diffusion-based generative framework, highlighting a growing interest in probabilistic modelling for robust AVSE.



Transformer-based audio-visual speech enhancement (AVSE) models have attracted significant attention in recent years owing to their exceptional capability to capture both local and global dependencies across modalities. Dual-transformer architectures \cite{wahab2024multi, afouras2018conversation} epitomize this approach by independently processing audio and visual streams, followed by alignment via self-supervised learning mechanisms. For example, DCUC-Net \cite{ahmed2023deep} extends the deep complex U-Net by incorporating Conformer blocks, which adeptly fuse convolutional and self-attention operations to facilitate more robust cross-modal integration. Moreover, iterative refinement techniques grounded in transformer frameworks \cite{nazemi2024iterative} have shown substantial gains in speech quality, especially under acoustically adverse conditions.

Despite these advancements, current AVSE architectures face several critical limitations. A predominant issue is their dependence on spectrogram-based inputs, as seen in models like AVDCNN \cite{Gabbay2017visual}, DCCRN \cite{hu2020dccrn}, and VSEGAN \cite{xu2022vsegan}, which can compromise fine temporal resolution and phase fidelity, ultimately affecting the precision of speech reconstruction. Furthermore, architectures such as AVDCNN \cite{Gabbay2017visual} rely on shallow or unidirectional fusion mechanisms, which constrain deep audio-visual integration and underutilize modality-specific correlations. Finally, transformer-dense models like AV-HuBERT \cite{Shi2022avhubert} often incur significant computational and memory costs, posing challenges for real-time deployment unless complemented by efficient model compression, hardware acceleration, or lightweight architectural innovations.

Motivated by the aforementioned limitations, we propose \texttt{AUREXA-SE}, an end-to-end audio-visual speech enhancement framework meticulously designed to address these challenges holistically. The architecture of \texttt{AUREXA-SE} is guided by four key design principles, each offering direct solutions to critical shortcomings in prior works:
\begin{itemize}
    \item \texttt{AUREXA-SE} employs a U-Net-like 1D convolutional audio encoder that directly operates on raw noisy waveforms. This design circumvents the limitations of spectrogram-based inputs and preserves fine-grained temporal details critical for accurate speech reconstruction.

    \item To address the constraints of shallow or unidirectional fusion strategies, \texttt{AUREXA-SE} introduces a novel bidirectional cross-attention mechanism. This iterative module enables deep, mutual contextualization between audio and visual modalities, fostering richer cross-modal integration.
    
    \item To ensure efficiency without sacrificing expressiveness, \texttt{AUREXA-SE} incorporates lightweight yet powerful components: a Swin Transformer V2 for hierarchical and efficient visual encoding, and a Squeezeformer module to model temporal dynamics with minimal computational overhead.
    

     \item For high-quality, perceptually faithful speech reconstruction, the fused embedding, refined through bidirectional attention, is passed through a U-Net–style decoder that directly synthesizes clean waveforms.
     
\end{itemize}

Building upon these innovations, \texttt{AUREXA-SE} seamlessly integrates state-of-the-art encoders with a robust bidirectional fusion strategy to effectively extract and align complementary cues from both modalities. This comprehensive design enables superior speech enhancement across a wide range of noisy environments, addressing the key shortcomings of prior architectures \cite{luo2019conv, liu2022swin, lin2022cat, kim2022squeezeformer, stoller2018wave}. The framework is also deeply inspired by foundational contributions in the field of audio-visual speech enhancement \cite{ michelsanti2021overview, yang2022audio}, grounding its innovations in both theoretical insight and empirical rigor.

\section{Methodology}
In this section, we delve deeper into the methodology and architectural design of \texttt{AUREXA-SE}, detailing how each component contributes to effective and high-quality AVSE in challenging environments.

\begin{figure}[htbp]
    \centering
    \includegraphics[width=0.35\textwidth]{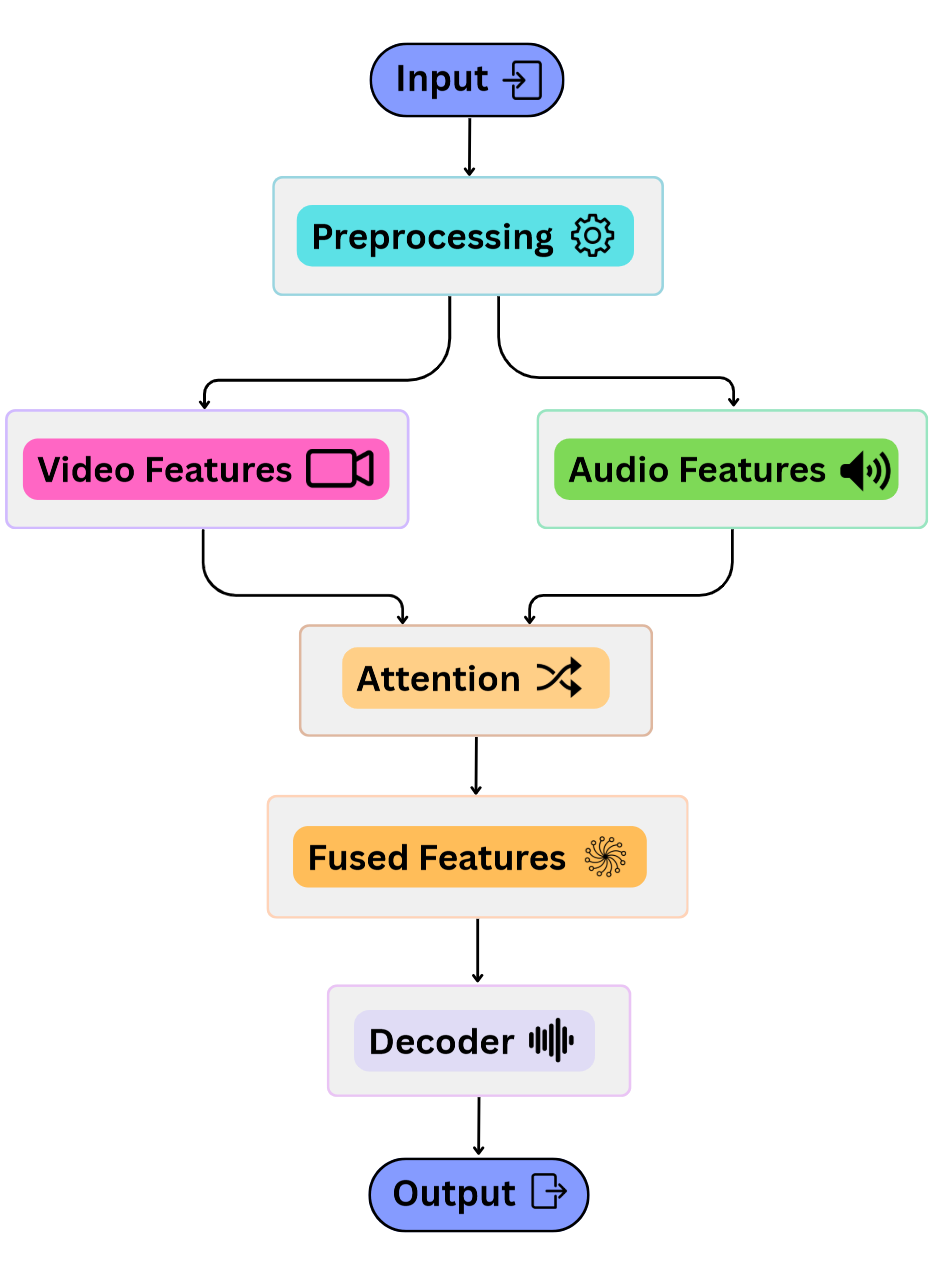}
    \caption{Architecture of proposed \texttt{AUREXA-SE}  framework}
    \label{fig:fig-architecture}
\end{figure}
\vspace{0.1cm}
\begin{figure*}[htbp]
    \centering
    \includegraphics[width=0.75\textwidth]{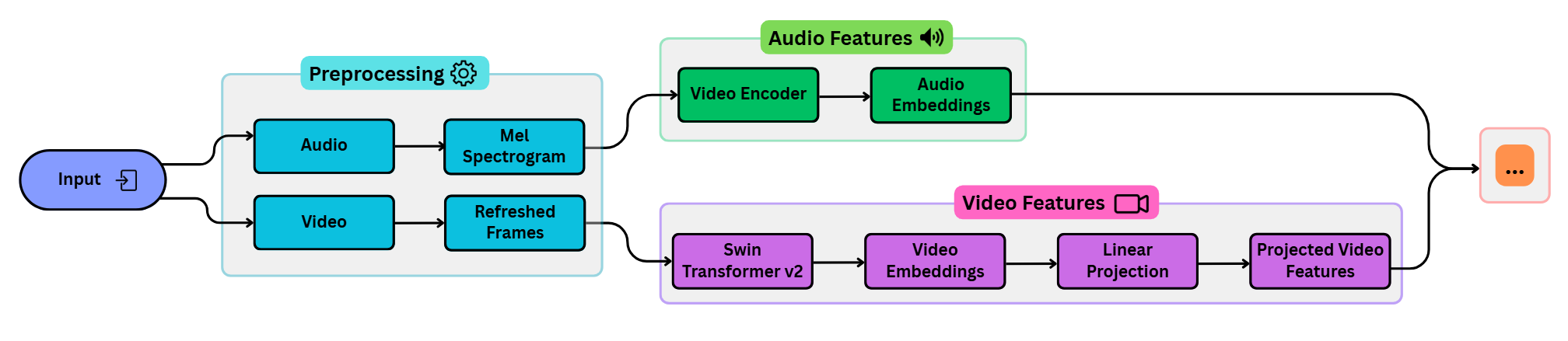}
    \caption{Detailed blueprint of data processing pipeline}
    \label{fig:data preprocess-architecture}
\end{figure*}

\begin{figure*}[htbp]
    \centering
    \includegraphics[width=0.75\textwidth]{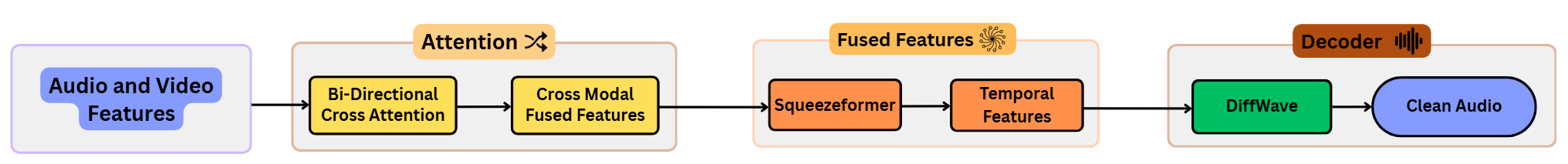}
    \caption{Detailed blueprint of attention and decoder mechanism}
    \label{fig:attention,decoder -architecture}
\end{figure*}

\begin{figure}[htbp]
    \centering
    \includegraphics[width=0.22\textwidth]{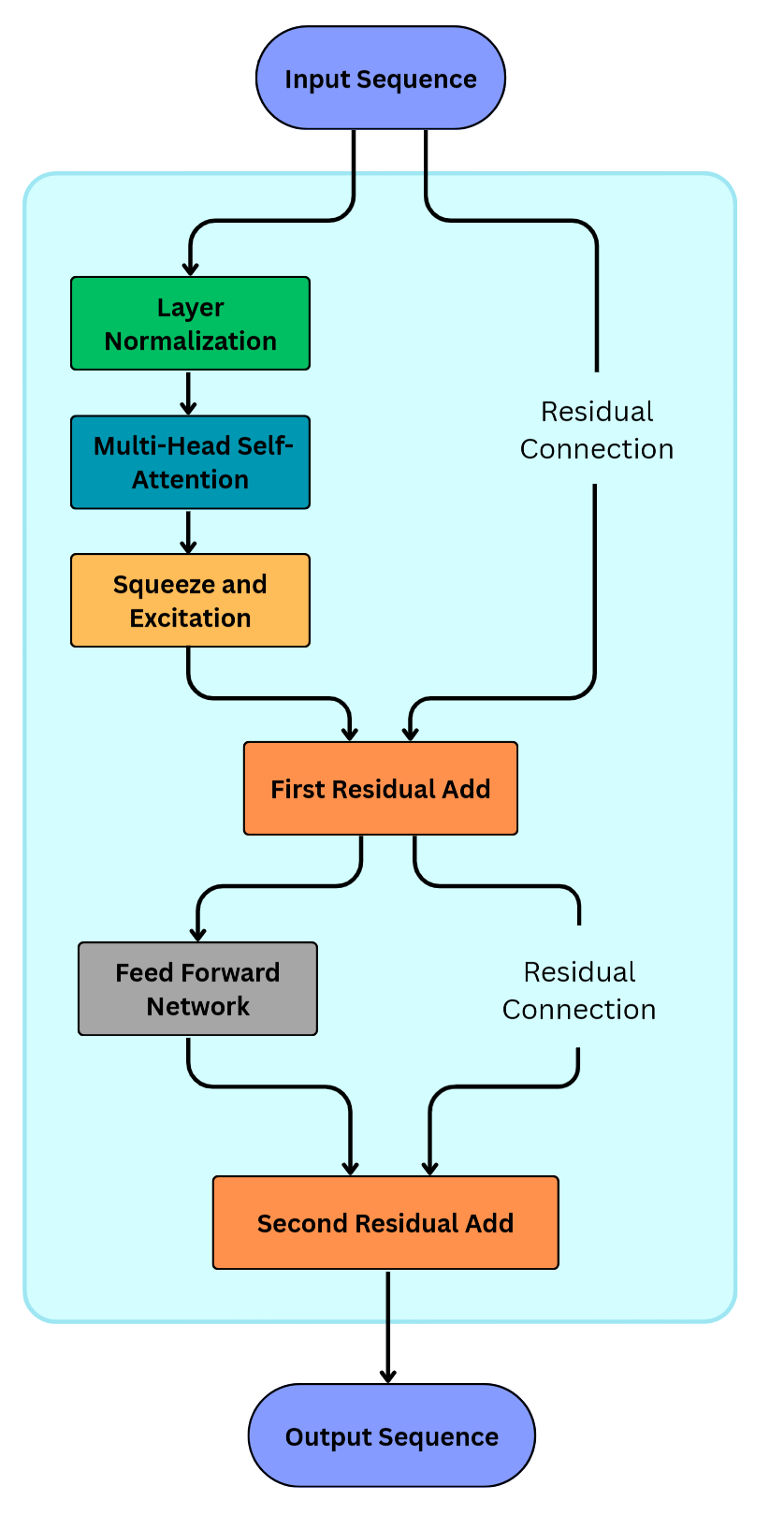}
    \caption{Workflow of a Squeezeformer block}
    \label{fig:squeezeformer-architecture}
\end{figure}

\subsection{Overview}
\texttt{AUREXA-SE} presents a novel bi-modal approach to speech enhancement, utilising both monoaural audio and visual information. Its architecture integrates a U-Net-based raw waveform audio encoder with a Swin Transformer V2 visual encoder, allowing the extraction of rich temporal and spatial features relevant to each modality \cite{stoller2018wave, liu2022swin}.The full system pipeline is depicted in Figure~\ref{fig:fig-architecture}. To achieve deep contextualisation between these diverse data streams, the system employs a bi-directional cross-attention mechanism for robust fusion. This is followed by a stage of temporal modelling, implemented through cascaded Squeezeformer blocks \cite{lin2022cat, kim2022squeezeformer}. The final component is a U-Net-inspired waveform decoder, incorporating skip connections, which upsamples and reconstructs the clean speech signal \cite{ronneberger2015u}. By drawing on both audio and visual information effectively, \texttt{AUREXA-SE} is designed to provide superior speech enhancement performance, even amidst noisy settings.

\subsection{Audio Encoder}
The audio encoder transforms raw, noisy audio into robust latent representations suitable for cross-modal fusion \cite{luo2019conv}. For stereo or multi-channel inputs, channels are averaged to produce a mono signal. The architecture follows a U-Net-inspired 1D convolutional design \cite{stoller2018wave} with 4 sequential downsampling blocks. Each block reduces the temporal resolution using a 1D convolution (kernel size 4, stride 2), followed by batch normalization and \texttt{ReLU} activation to stabilize and non-linearize the feature maps.

This hierarchical structure enables the model to extract multi-scale temporal features, allowing it to retain essential speech patterns even under severe noise. After downsampling, a 1×1 convolution projects the features into a fixed-dimensional latent space. The final output is reshaped to \texttt{[Batch, Time, Feature\_Dim]}, forming a compact, noise-resilient audio embedding aligned for fusion with visual features. This component is mentioned in Figure~\ref{fig:data preprocess-architecture}.

\subsection{Video Encoder}
Visual cues such as lip movements and facial expressions provide context for SE, especially under noisy conditions. As illustrated in Figure~\ref{fig:data preprocess-architecture}, each input clip consists of 75 RGB frames resized to 112×112 pixels. These frames undergo preprocessing, including dimension reordering to \texttt{[Batch × Time, Channels, Height, Width]} and pixel value normalization through clamping.

Each frame is independently encoded using a Swin Transformer V2 \cite{liu2022swin}, a hierarchical vision transformer that models multi-scale spatial features via local and shifted window-based self-attention. After flattening the frames along the temporal dimension to form a tensor of shape \texttt{[Batch × Time, C, H, W]}, the resulting sequence is processed using the Swin Transformer V2 encoder. The output hidden states are globally pooled across tokens and projected to a fixed 512-dimensional embedding via a linear layer.

These per-frame embeddings are reshaped back to a temporal sequence with shape \texttt{[Batch, Time, Feature\_Dim]} and optionally undergo further projection, normalization, and clamping. This process yields rich, temporally aligned visual features optimized for subsequent cross-modal fusion with audio cues.

\subsection{Cross-Modal Fusion via Bi-directional cross-attention}
The architecture employs a sophisticated bi-directional cross-attention \cite{vaswani2017attention, lin2022cat} mechanism, as shown in Figure~\ref{fig:attention,decoder -architecture}, to deeply integrate audio and visual features, enabling mutual contextualisation. Before fusion, video features are temporally aligned with audio features via linear interpolation if their sequence lengths differ.

This fusion occurs through an iterative process in which a dedicated \texttt{nn.MultiheadAttention} layer allows audio features to query video features (audio-to-video attention) and, simultaneously, video features to query audio features (video-to-audio attention). This bi-directional interaction ensures each modality is updated with relevant context from the other. Residual connections and \texttt{nn.LayerNorm} are applied after each attention operation to stabilise learning.

This iterative refinement yields the fused audio and video representations, which are then averaged to obtain a unified latent representation. This combined feature undergoes clamping and prepares the robust, fused embeddings for subsequent processing.

\subsection{Temporal Modeling}
The cross-modal fusion process yields enhanced audio-visual embeddings that require temporal modelling to capture sequential speech patterns. The temporal modelling component consists of stacked Squeezeformer \cite{kim2022squeezeformer} blocks applied to the fused feature sequence$\ F \in \mathbb{R}^{B \times T \times D}$. The overall architecture of this component is visualized in Figure~\ref{fig:squeezeformer-architecture}.

\subsubsection{Squeezeformer Architecture}
Each block combines a squeeze operation for temporal downsampling, multi-head self-attention for global dependencies, depth-wise separable convolutions for local patterns, and position-wise feed-forward networks with residual connections. The Squeezeformer architecture reduces computational complexity from $O(T^2)$ to $O(T\ log( T))$ through its squeeze operation, making it suitable for processing raw audio waveforms where T = 37,830 samples (34,524 for the training set and 3,306 for validation) corresponds to 3 seconds of 16 kHz audio.

\subsubsection{Cross-Modal Temporal Dependencies}
Operating on post-fusion embeddings enables the model to learn joint sequential patterns, capturing synchronisation between visual speech cues and acoustic events. The temporal model generates features that maintain their original sequence lengths while embedding a comprehensive temporal context essential for the reconstruction of waveforms. These features are then used as conditioning for the diffusion decoder.

\subsection{Decoder}
The decoder reconstructs clean speech waveforms from fused audio-visual features using a U-Net–inspired architecture \cite{stoller2018wave}, also illustrated in Figure~\ref{fig:attention,decoder -architecture}. It begins with a linear projection to match encoder skip connection channels, followed by a series of upsampling blocks that progressively double the temporal resolution. Each block uses linear layers with \texttt{LayerNorm} and \texttt{ReLU} to ensure stable training.

Skip connections \cite{ronneberger2015u} from the audio encoder are incorporated at each stage, with alignment handled via interpolation and padding when needed. A final linear layer with \texttt{Tanh} activation generates the waveform output in the 
clamped range. This design preserves fine-grained audio details, improves gradient flow, and enables high-fidelity waveform reconstruction.

\subsection{Loss Function and Evaluation Metrics}
The model is optimized using the Mean Squared Error (MSE) loss, a fundamental and widely used objective function that encourages similarity between the predicted and target waveforms. For validation, we employ perceptual and intelligibility-based metrics, namely Perceptual Evaluation of Speech Quality (PESQ), Scale-Invariant Signal-to-Noise Ratio (SI-SNR), and Short-Time Objective Intelligibility (STOI). These metrics quantitatively assess the fidelity of the predicted speech by comparing it to the corresponding clean reference, guiding the model to reduce reconstruction errors and improve perceptual quality. PESQ scores range from -0.5 to 4.5 and reflect perceptual quality, while STOI scores from 0 to 1 indicate intelligibility. SI-SNR (or SISDR) quantifies distortion, with higher values denoting better signal fidelity.

\section{Experiments}
This section outlines the dataset used, describes the experimental setup, and provides a comprehensive discussion of the evaluation results.
\subsection{Dataset Description}
The AVSE-4 dataset used in our study is publicly available on GitHub \cite{avse4data2025}
, which consists of audio-visual scenes combining speech and noise under both synthetic and real-world acoustic conditions. Each scene features a target speaker and up to three interferers drawn from competing speakers, non-speech noises (e.g., domestic appliances, human sounds), and music tracks from MedleyDB. Scene construction follows the clarity challenge methodology, using a speech-frequency-weighted SNR ranging from -10 dB to +10 dB during training and -18 dB to +6.55 dB in evaluation.

The training set includes 34,524 scenes (113 hours) with 605 target speakers and 15 noise types, while the development set contains 3,306 scenes (9 hours) with 85 target speakers. Each scene provides a silent video, mixed mono audio, and isolated audio tracks for the target and interferers. All audio is 16 kHz, 16-bit, and the dataset includes facial landmarks and embeddings (e.g., FaceNet, Facemesh) to support visual modelling. The out-of-domain set includes real conversational speech in acoustically controlled settings.

\subsection{Experimental Setup}

\begin{figure}[htbp]
    \centering
    \includegraphics[width=0.4\textwidth]{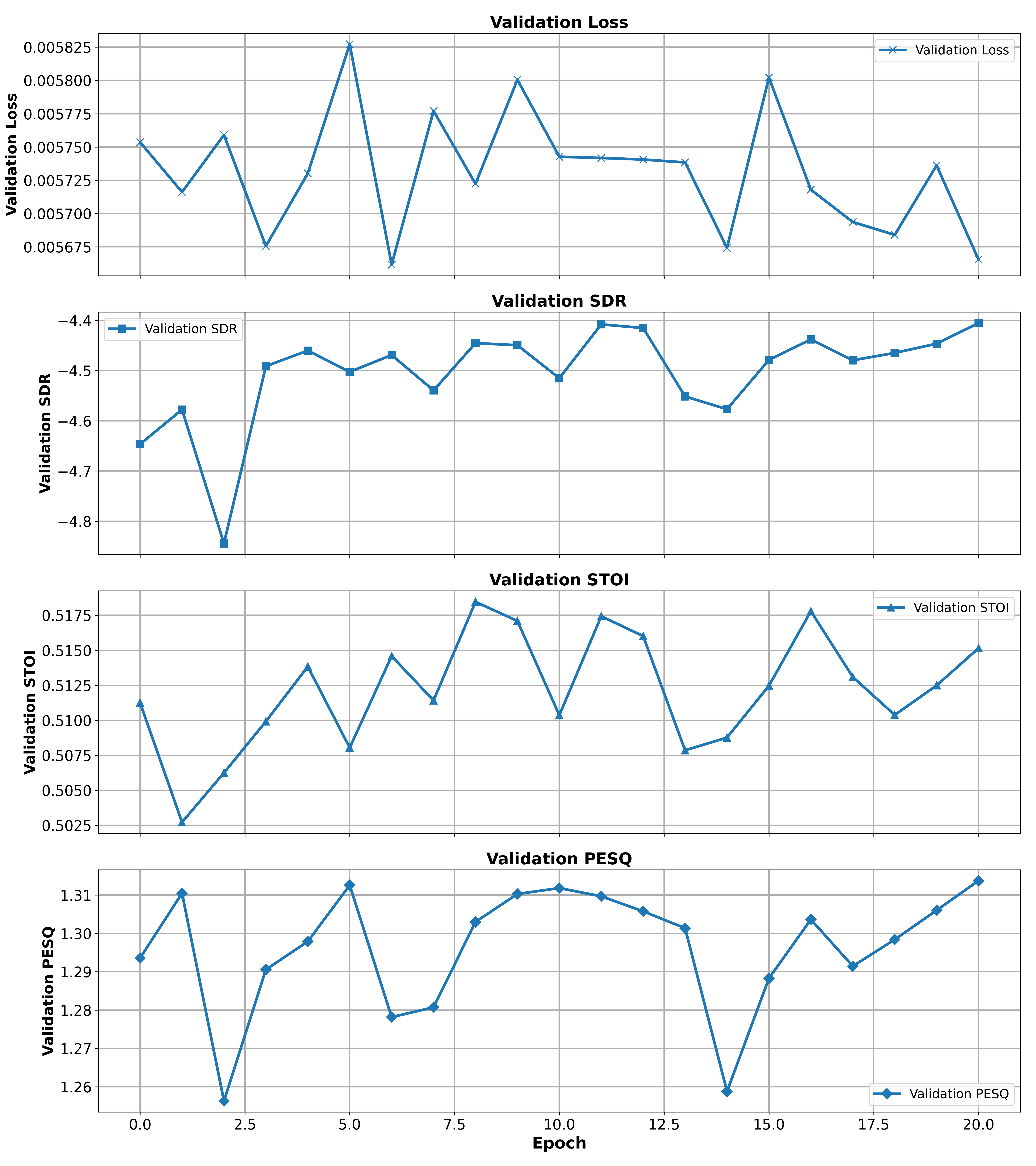}
    \caption{Validation performance of the \texttt{AUREXA-SE} model across 20 epochs. The graph illustrates trends in SDR, PESQ, STOI, and validation loss.}
    \label{fig:aurexa-metrics}
\end{figure}

We use \texttt{AVSE4Dataset} and \texttt{AVSE4DataModule} to prepare training, validation, and test sets, with each sample clipped or padded to 3 seconds (75 video frames at 25 FPS and 48,000 audio samples at 16 kHz). Experiments were conducted on an NVIDIA RTX A4500 GPU with 46 GB RAM. The proposed \texttt{AUREXA-SE} model consists of 54.2 M trainable parameters, resulting in an estimated size of 217.859 MB. Training spanned 20 epochs and a total of 344,680 steps.

During preprocessing, videos are resized to 112×112, and audio is normalised. Inputs are clipped or padded for consistency, with audio processed in mono or stereo and visuals standardised as RGB.





\begin{table}[htbp]
\centering
\caption{Comparison of PESQ, STOI, and SISDR across models}
\begin{tabular}{lccc}
\hline
\textbf{Model} & \textbf{PESQ} & \textbf{STOI} & \textbf{SISDR} \\
\hline
Noisy Input    & 1.171 & 0.459 & -5.847 \\
Baseline    & 1.227 & 0.487 & -5.125 \\
\texttt{AUREXA-SE}         & \textbf{1.325} & \textbf{0.514} & \textbf{-4.312} \\
\hline
\label{table:table-eval}
\end{tabular}
\end{table}
\vspace{-2mm}
\subsection{Evaluation Results}
During the evaluation, three types of audio samples were considered. First, we used the noisy speech directly from the AVSE4 testing dataset. This unprocessed audio served as the input to all models and acted as the standard for comparison. Second, we applied the COG-MHEAR AVSE Challenge 2024 baseline model to enhance this noisy audio. Third, we used our proposed \texttt{AUREXA-SE} model to perform speech enhancement on the same input. Each of these versions was evaluated using three standard objective metrics: PESQ, STOI, and SI-SDR. The final scores obtained by all models are presented in Table~\ref{table:table-eval}.

\begin{figure}[htbp]
    \centering
    \includegraphics[width=0.4\textwidth]{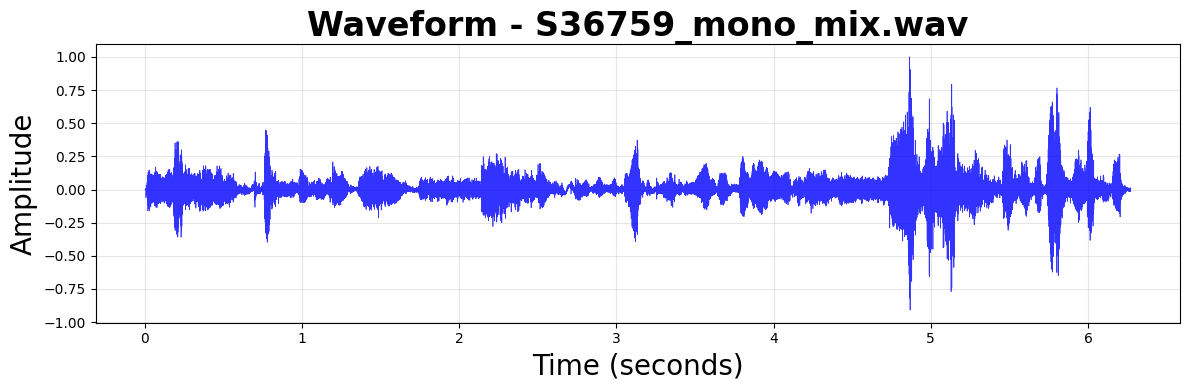}
    \includegraphics[width=0.4\textwidth]{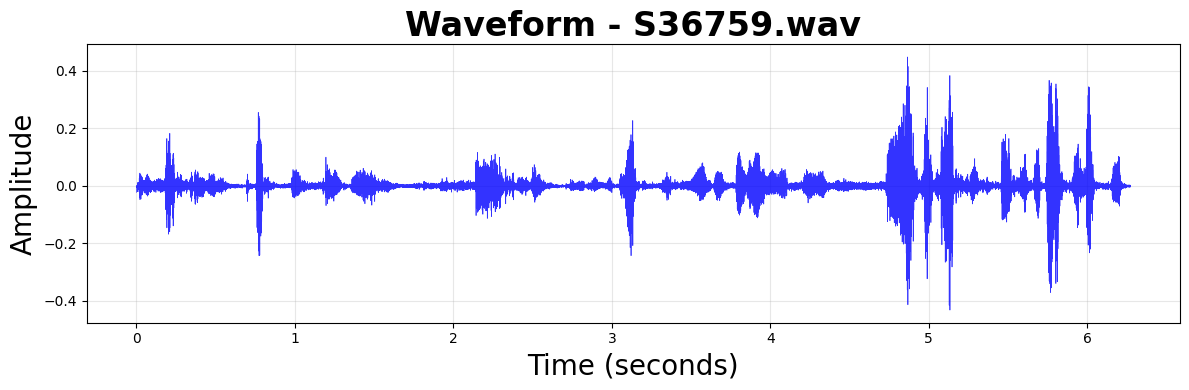}
    \caption{Waveform comparison for a sample from the test set. (Top) The original noisy audio waveform. (Bottom) The enhanced waveform after being processed.}
    \label{fig:waveform_comparison}
\end{figure}

\begin{figure}[htbp]
    \centering
    \includegraphics[width=0.28\textwidth]{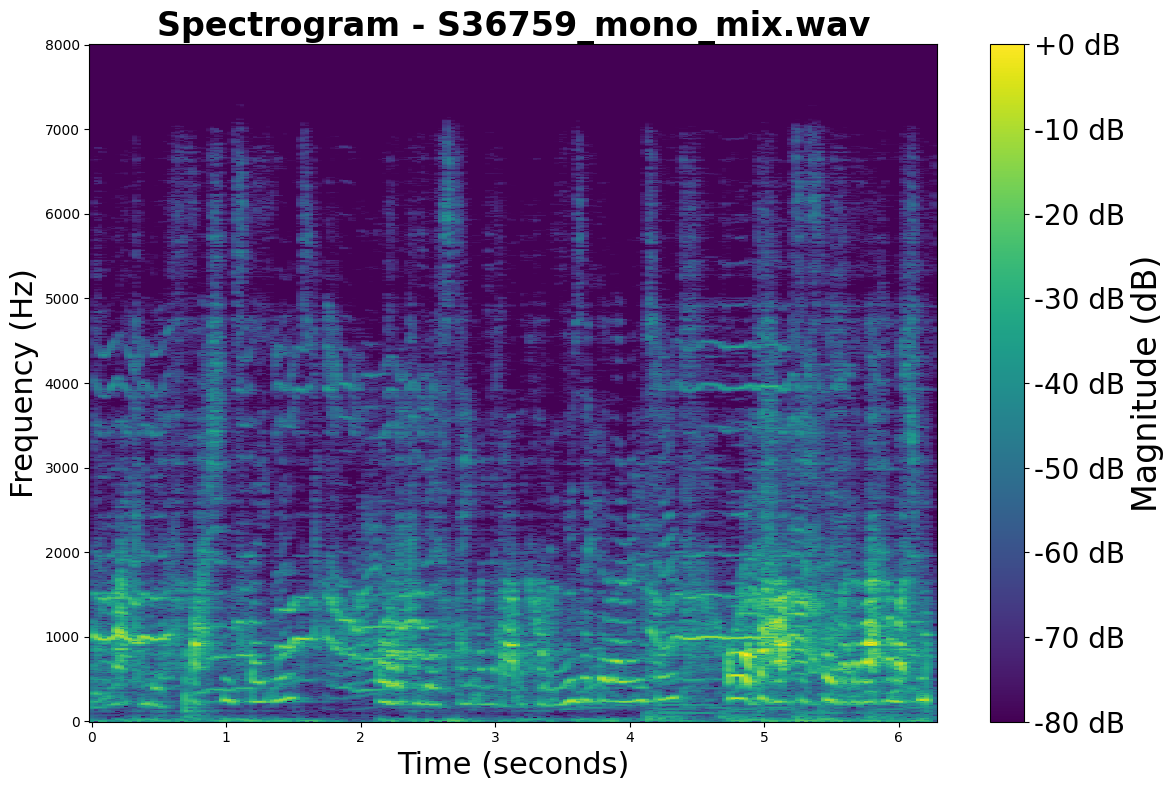}
    \includegraphics[width=0.28\textwidth]{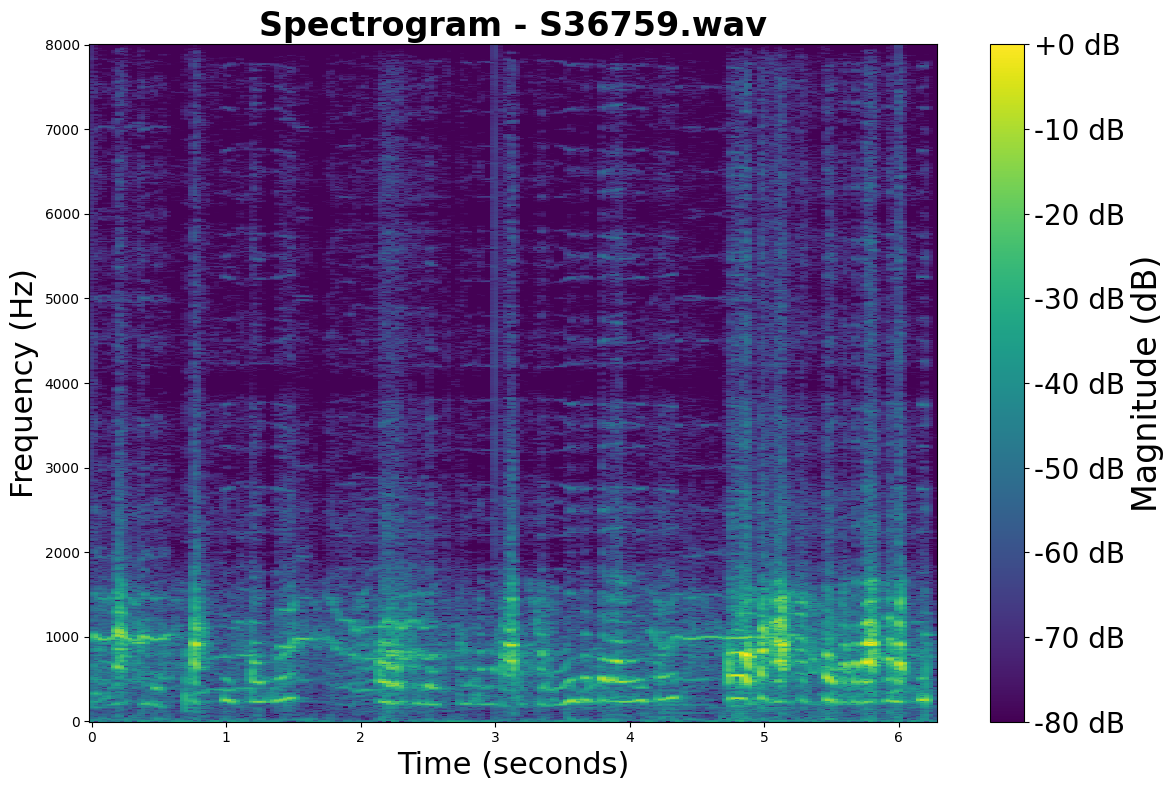}
    \caption{Spectrogram visualization of speech enhancement. (Top) The spectrogram of the noisy input, where background noise artifacts obscure the speech harmonics. (Bottom) The spectrogram of the output from \texttt{AUREXA-SE}.}
    \label{fig:spectrogram_visualization}
\end{figure}

Compared to the noisy input, the baseline model showed notable improvements in both quality and intelligibility. The PESQ score increased from 1.171 to 1.227, and the STOI improved from 0.459 to 0.487. The proposed \texttt{AUREXA-SE} model achieved the best results across all evaluation criteria, with a PESQ score of 1.325, a STOI of 0.514, and a SI-SDR of -4.312 dB. This represents a relative gain over the baseline of +0.098 in PESQ, +0.027 in STOI, and +0.813 dB in SI-SDR. Figure~\ref{fig:aurexa-metrics} further illustrates the validation trends across 20 epochs, showing consistent improvements in PESQ, STOI, SI-SDR, and loss over time. Overall, the evaluation confirms that \texttt{AUREXA-SE} surpasses both the unprocessed noisy audio and the AVSE baseline across all standard metrics.

\noindent \textbf{Computational Cost:} Regarding the computational cost, the superior performance of \texttt{AUREXA-SE}, which surpasses the baseline, was achieved in just 50 hours of total training. This 20-epoch training period (348,660 steps) represents a highly effective path to state-of-the-art results, especially when compared to the week-long training cycles often required for other advanced architectures. While our model introduces a trade-off in inference speed, requiring 40 minutes versus the baseline's 25, its ability to deliver top-tier enhancement quality from a modest 50-hour training investment underscores its potent and well-balanced design.
\vspace{-2mm}
\section{Conclusion and Future Work}
In this work, we proposed \texttt{AUREXA-SE}, a unified architecture for audio-visual speech enhancement (AVSE) that addresses the limitations of audio-only systems in challenging acoustic conditions. By leveraging complementary cues from both modalities, \texttt{AUREXA-SE} captures richer context for more robust speech recovery. The architecture integrates a Swin Transformer V2 \cite{liu2022swin} for spatial visual encoding, a U-Net–based raw waveform encoder \cite{stoller2018wave} for acoustic detail, and a bidirectional cross-attention mechanism \cite{lin2022cat} for deep fusion. The fused features are temporally modeled using Squeezeformer \cite{kim2022squeezeformer} and decoded via a U-Net-inspired waveform decoder \cite{ronneberger2015u} to reconstruct high-fidelity speech. Experimental results on the AVSE4 benchmark show that \texttt{AUREXA-SE} effectively models cross-modal dependencies and consistently outperforms noisy baselines, demonstrating its potential for real-world deployment. Ongoing future work aims to address current limitations of the proposed model, specifically its robustness to unseen noise types, and extend the architecture to support multi-speaker separation. 

Looking ahead, we plan to enhance \texttt{AUREXA-SE} with real-time capabilities, improved fusion strategies, and robust visual encoding to better handle real-world challenges. A comprehensive comparative evaluation will further validate its generalization and effectiveness in diverse noise conditions.

\section{Acknowledgement} Prof Hussain acknowledge the support of the UK Engineering and Physical Sciences Research Council (EPSRC) Grant Ref. EP/T021063/1 (COG-MHEAR) and EP/T024917/1 (NATGEN). The work
of M. Sajid is supported by the Council of Scientific and Industrial Research (CSIR), New Delhi for providing fellowship under the under Grant 09/1022(13847)/2022-EMR-I.

\bibliographystyle{IEEEtran}
\bibliography{mybib}
\end{document}